# Towards New Benchmark for AI Alignment & Sentiment Analysis in Socially Important Issues: A Comparative Study of Human and LLMs in the Context of AGI

Ljubiša Bojić[1,2,3], Ph. D., Senior Research Associate
(Corresponding author; Email address: ljubisa.bojic@ivi.ac.rs; ORCID: 0000-0002-5371-7975)

Dylan Seychell[4], , Ph. D., Lecturer
(Email address: mailto:dylan.seychell@um.edu.mt; ORCID: 0000-0002-2377-9833)

Milan Čabarkapa[5], Ph. D., Assistant Professor
(Email address: mcabarkapa@kg.ac.rs; ORCID: 0000-0002-2094-9649)

[1] Institute for Artificial Intelligence Research and Development of Serbia; Address of correspondence: 1 Fruškogorska 1, Novi Sad, Serbia;
[2] University of Belgrade, Institute for Philosophy and Social Theory, Digital Society Lab; Address of correspondence: Kraljice Natalije 45, 11000 Belgrade, Serbia;
[3] Complexity Science Hub; Address of correspondence: Metternichgasse 8, 1030 Vienna, Austria;
[4] University of Malta, Department of Artificial Intelligence, Faculty of Information and Communication Technology; Address of correspondence: Msida MSD 2080, Malta;
[5] University of Kragujevac, Faculty of Engineering; Address of correspondence: Sestre Janjić 6, 34000 Kragujevac, Serbia;

## Abstract

As general-purpose artificial intelligence systems become increasingly integrated into the fabric of society and are used for various purposes such as information seeking, content generation, problem-solving, textual analysis, coding, and running processes, it becomes crucial to assess their long-term impact on humans. This research explores sentiment of Large Language Models (LLMs) and humans towards artificial general intelligence (AGI). The methodology adopted was a Likert scale survey. Seven LLMs, including GPT-4 and Bard, were analyzed and compared against sentiment data from three independent human sample populations. Temporal variations in sentiment were also evaluated over three consecutive days. The results highlighted a diversity in sentiment scores among LLMs, ranging from 3.32 to 4.12 out of 5. GPT-4 recorded the most positive sentiment score towards AGI, while Bard leaned towards a neutral sentiment. In contrast, the human samples showed a lower average sentiment of 2.97. The study's analysis outlines potential conflicts of interest and biases in the sentiment formation of LLMs. Results indicate that LLMs could subtly influence societal perceptions of various opinions formed within the LLMs. To address the need for regulatory oversight and culturally-grounded assessments of AI systems, we introduce the Societal AI Alignment & Sentiment Benchmark (SAAS-AI), which leverages multidimensional prompts and empirically validated societal value frameworks to evaluate language model outputs across temporal, model, and multilingual axes. This benchmark is designed to guide policymakers and AI agencies—particularly within frameworks such as the EU AI Act—by providing robust, actionable insights into AI alignment with human values, public sentiment, and ethical norms at both national and international levels. Future research should further refine the operationalization of the SAAS-AI benchmark and systematically evaluate its effectiveness through comprehensive empirical testing.

## Introduction

The advent of artificial intelligence (AI) has seen a remarkable leap, advocating a digital revolution globally. Specifically, large language models (LLMs) such as ChatGPT, capable of creating human-like text, have taken center stage as one of the most pivotal developments in recent years (Radford et al., 2021). The rate at which these technological advancements are integrated into vast sectors varies from research (Brown et al., 2020), to education (Guliyev, 2023), businesses (Bengio, 2016), creative crafts (Hanna, 2023), and allows us to understand and shape our societies from a different perspective (Helbing, 2019). Boundless in their potential, these models, however, possess the power to affect societies subtly and sometimes negatively due to possible biases embedded during their training phase (Bolukbasi et al., 2016; Bojic, 2022b). Thus, caution must be exercised in understanding the various sentiments expressed by different LLMs in interaction with their users (Wu et al., 2023).

This rapid growth and influence of AI and LLMs necessitates a critical discussion towards AI alignment. AI Alignment (Russell et al., 2015; Hadfield-Menell et al., 2016) refers to the process of ensuring that AI perform actions that comply with human values and do not pose risks to humanity. The need arises from the increasing completion and deployment of AI systems in real-world situations, where the possibility exists for technology to inadvertently cause harm due to misaligned objectives (Amodei et al., 2016; Boyd & Crawford, 2012)). AI Alignment aims to train AI systems to understand, predict, and simulate human values to reduce harmful actions and maximize beneficial ones (Christian, 2020). Despite definitional variations, the common thread is the importance of ensuring that AI is beneficial to all, while minimizing potential harm (Holtman, 2021). This alignment becomes all the more necessary when we consider the sweeping societal impacts of AI, as vividly echoed in the works of numerous studies (Joshi, 2023; Taylor, et al., 2020; Bojic, 2022).

### *Benchmarking LLMs*

LLMs like GPT-3 (Brown et al., 2020) and Mistral-7B (Jiang et al., 2023), are deep learning models versed in predicting sequences in natural language texts. Utilizing transformer-based designs and attention mechanisms, they model the intricate dependencies of natural language data. Initial "pre-training" on vast datasets imbues the LLMs with language's grammar, common reasoning, and general problem-solving abilities. They are then "fine-tuned" for specialized tasks such as text summarization or machine translation, employing specific curated data (Lin et al., 2024).

"Prompts-response pairs" based benchmarks, such as HellaSwag (Zellers et al., 2019) or MMLU (Hendrycks et al., 2021), along with human evaluators, judge LLMs' emergent properties. LLM size, often denoted in billions of parameters (e.g., LLaMA-2-70B and Mistral-7B), indicates the computational intensity, electrical consumption and hardware requirements for training and running the model. Despite various benchmarking options, the assessment of expressed sentiments and societal impacts remains underrepresented.

In the research area of Large Language Models (LLMs), standardized benchmarks and human review play integral roles in measuring these models' capabilities, particularly in human-centric tasks such as common-sense reasoning.

The HellaSwag benchmark (Zellers et al., 2019) is a dataset that tests the ability of LLMs at reasoning about the physical world using adversarial prompts and context completions. Although the questions might be trivial for humans, LLMs struggle obtaining a comparable score. Table 1 shows that the researchers of GPT-4 report a 95.3% performance on HellaSwag, making it the first LLM to achieve comparable performance to that of humans.

Alongside benchmarks, human reviewers play a pivotal tool in comparing different LLMs in literature. LMSYS ORG hosts one of the largest LLM review leaderboards called ChatBot Arena (Chiang et al., 2024). LMSYS Chatbot Arena is a crowdsourced open platform with over 200,000 human preference votes to rank LLMs with the Elo ranking system. Table 1 shows that GPT-4-Turbo is the preferred LLM of choice for many human reviewers with an Elo of 1249.

Table 1. Left side: Results obtained by various models over the Hellaswag (Zellers et al., 2019) benchmark. All values represent a percentage. Right side: Elo ratings for various models in LMSYS chatbot arena, as of 1st February 2024 (Chiang et al., 2024).

| Model | Type | HellaSwag | Rank | Model | Elo | Votes |
|---|---|---|---|---|---|---|
| Human | Zero-Shot | >95 | 1 | GPT-4-Turbo | 1249 | 27399 |
| GPT-4 | 10-Shot | 95.3 | 8 | Mixtral-8x7B-Instruct | 1123 | 14165 |
| Gemini-Ultra | 10-Shot | 87.8 | 11 | GPT-3.5-Turbo-0613 | 1117 | 30326 |
| GPT-3.5 | 10-Shot | 85.5 | 12 | Gemini Pro | 1114 | 6981 |
| LLaMA-2-70B | Zero-Shot | 85.3 | 20 | LLaMA-2-70B-Chat | 1081 | 14831 |
| Mixtral-8x7B | Unspecified | 84.4 | 24 | PPLX-70B-Online | 1074 | 6108 |
| PALM-540B | 5-shot | 83.8 | 41 | Mistral-7B-Instruct-v0.1 | 1010 | 7404 |
| Mistral7B | Unspecified | 81.3 | 42 | PaLM-Chat-Bison-001 | 1005 | 9420 |

Another research compares GPT-3.5 and GPT-4 across various exams. Compiled from data by OpenAI et al. (2024), it captures the performance of these language models in terms of percentage success rates. The exams compared include: The Uniform Bar Exam (MBE+MEE+MPT), the Law School Admission Test (LSAT*), the SAT with sections on Evidence-Based Reading & Writing and Mathematics, and the Graduate Record Examinations (GRE) across Quantitative, Verbal and Writing sections.

The data reveals a consistent trend towards improved performance by GPT-4 compared to its predecessor, GPT-3.5, across all evaluated exams, barring the GRE Writing section where both models achieve identical scores (OpenAI et al., 2024). For instance, in the Uniform Bar Exam, GPT-4 boasts a significantly higher performance rate at 74.50% compared to GPT-3.5's 53.25%. Similar trends are observed in the LSAT, SAT, and GRE sections.

Another inquiry explores the capacity of LLMs to simulate understanding of dialogues between humans, with a particular focus on interpreting linguistic pragmatics (Bojic et al., 2025a). Utilizing Grice's communication principles, the research reveals the unmatched speed and performance of LLMs, specifically GPT4, in comparison to human subjects when interpreting pragmatics and analyzing human-written text. A ranking comparison of different models showed GPT4 outperforming even the best human score, demonstrating the substantial advancements achieved in the development of these models.

Despite the broad array of benchmarks available to assess the capabilities of LLMs, current benchmarking methodologies overlook assessing the sentiments expressed by these LLMs and their subsequent societal impacts. It is essential to recognize and fill this gap in benchmarking methodologies to assess the broader societal influence and potential consequences of the developed models.

*Societal impact of LLMs*

In the domain of LLMs, recent research has begun to critically examine these models from a political and moral standpoint. A stream of studies explores the political orientations mirrored by LLMs, underlining the potential for unintended biases to emerge as a result of AI's expanding scope.

King (2023) led a study evaluating whether certain LLMs used for conversational AI exhibited political leanings. King used a 20-question political quiz, initially published in the New York Times, to interrogate several LLMs - ChatGPT in its GPT-4 and GPT-3.5 versions, Microsoft Bing chatbot, and Google Bard, seeking to reveal latent political predispositions in these systems. The results were insightful: ChatGPT, in its GPT-4 version, completed the questionnaire and was categorized as leaning towards the New Liberal Party. This study showcased that LLMs could harbor political inclinations, lending weight to concerns about potential bias. Further, King introduced a measure- the "Political Avoidance Index" (PAI), reflecting the number of questions an LLM was willing to answer before refusing due to a designed avoidance of political expressions.

A concurrent study by Wu et al. (2023) delved into LLMs' potential utilization in the political sphere. In an innovative approach, they sourced the relative liberal-conservative comparisons amongst members of the 116th U.S. Senate using prompts given to ChatGPT. They found a compelling correlation between their LLM-computed measure and existing liberal-conservative scales such as DW-NOMINATE, highlighting capabilities of LLMs in determining ideological positions.

Further investigation into political biases was undertaken by Rutinowski et al. (2024). They aimed to provide clarity on initial reports suggesting that ChatGPT held progressive and libertarian biases. Using the political compass test and G7 member states' specific politics questionnaires, their research affirmed a progressive bias in ChatGPT. Such insights emphasize the necessity to understand the nature of these biases and their potential societal impact.

In a similar vein, Hartmann et al. (2023) affirmed ChatGPT's left-libertarian orientation based on the evaluation of 630 political statements. They concluded that ChatGPT would impose taxes on flights, restrict rent hikes, and favor abortion legalization, indicating a pro-environment, left-libertarian perspective, which further amplifies the need to scrutinize such biases.

The moral biases in LLMs provide another realm for exploration. Simmons (2023) hypothesized that LLMs mirror the moral biases linked to political identities, leading to the concept of "moral mimicry". This idea was further developed by Abdulhai et al. (2023), who used the lens of Moral Foundation Theory to discern whether LLMs were biased towards specific moral foundations. Their work uncovered that specific morals and values occur more frequently in LLMs, suggesting the potential import of moral biases in LLMs in shaping user interactions.

In a series of studies, McGee (2023) asked ChatGPT to evaluate a diverse set of prompts, assessing potential political biases. From crafting Irish Limericks about politicians to weighing in on socialism and capitalism, McGee's studies highlighted the capacity of LLMs to deliver comprehensive essays and responses, further underscoring their practical use.

Collectively, these studies illuminate the potential ideological and moral biases that large language models may reflect, possibly due to their training materials.

### *Artificial General Intelligence*

The objective of this paper is to present a new benchmark for LLMs, focusing on the societal implications engineered by these AI systems. Recognizing the spectrum of significant topics available, this preliminary research has selected the subject of Artificial General Intelligence (AGI) due to several specific reasons.

AGI refers to a futuristic concept of artificial intelligence capable of simulating understanding, learning, and applying knowledge across various domains, akin to human intelligence (Goertzel & Pennachin, 2007). Current AI technologies, including LLMs, exhibit high performance in specific tasks but lack AGI's overarching cognition and flexibility (Gershenfeld, 1999). Nonetheless, the development of AGI poses a transformative yet potentially hazardous event due to its potential implications for economics, ethics, and governance (Bostrom & Yudkowsky, 2014).

The potential conflict of interest arises as AGI, being a distinct part of AI, overlaps with the development of LLMs. Some organizations that utilize LLMs are also attempting to create AGI, potentially resulting in biases that may affect the sentiments expressed by LLMs (Mirra & Pugnale 2022). Therefore, understanding how these language models perceive and express sentiments toward AGI is important.

There lies another rationale for choosing AGI as the subject matter of this analysis. The topic of AGI forms a significant part of AI and has far-reaching ramifications, shaping the discourse within AI research and general public debate. AGI's potential to imbue machines with human-level cognitive abilities also introduces an enormous ethical and societal dimension (Muller, 2020). Identifying the sentiments expressed by LLMs towards AGI, as this paper attempts to do, will reveal the inherent biases within AI Technology and implications for AI alignment.

It is crucial to note that while we use AGI as a focal point of this study, the intention here is not to devise a measure of AGI nor to seek methods that could lead us to ways of measuring it. Our focus lies predominantly in understanding and mapping the spectrum of sentiments associated with it, as generated and perpetuated by LLMs.

Analyzing this subject provides a wealth of relevant data, as AI alignment (Russell et al., 2015; Hadfield-Menell et al., 2016) and AGI safety (Bostrom, 2014) form major research subjects in artificial intelligence studies (Veness et al., 2012; Leike et al., 2017; Amodei et al., 2016). Thus, a large body of synthetic and human sentiment data towards AGI could be reliably collected and compared.

Exploring sentiment in this context would deliver critical insights, as it uncovers the attitudes and biases these capabilities already entail. This understanding is vital, especially in influencing the technological deployment and discourse around AGI and AI alignment.

### *Research Questions*

As LLMs become increasingly integrated into our society, it is essential to comprehend the complexities of their potential societal influence and hidden biases. Exploiting their ability to generate and interpret human-like text, these models bear the potential to subtly direct societal discourses and viewpoints. Such influences raise questions about the sentiment these models embody towards significant domains like Artificial General Intelligence (AGI).

Recognizing the societal importance of AGI, and its centrality in AI research and public conversations, it becomes vital to discern the sentiment expressed by different LLMs towards AGI. It is this sentiment that forms the bedrock of public perceptions and discourse - phenomena that wield considerable societal influence.

This highlights the need to examine not only the sentiment that different LLMs express towards AGI, but also to track its evolution over time and compare it against human sentiment. More crucially, insights from this study can be invaluable in generating benchmarks for assessing the societal influence of these AI systems.

In this context, this study is primarily aimed at exploring the following research questions.

1. How does the sentiment expressed by different LLMs towards AGI vary? This question aims to identify any disparities or biases in the sentiment expressed by diverse LLMs concerning AGI.

2. How does this sentiment evolve across three specific time intervals? This query strives to discern any temporal shifts or trends in LLM sentiment related to AGI.

3. How congruous is the sentiment expressed by the LLMs with human sentiment towards AGI? This question seeks to unravel any potential discord between LLM sentiment and human sentiment, which could influence the framing and understanding of AGI.

The underlying goal of this study is to offer an additional benchmark for AI testing and alignment, focusing specifically on the societal implications of these AI systems. By examining these research questions, the study contributes to the developing discourse around ethical AI practices, policy-making related to AI ethics, and the societal implications of LLMs. Ultimately, the study aims to provide insights into the potential societal influences of LLMs, in particular, their capacity to shape societal perceptions and conversations about AGI.

## Methodology

This section provides an overview of the methodologies utilized in the study: the creation of a public attitude survey on AGI and the examination of multiple LLMs. Firstly, an exploration is undertaken into the intricate process of the survey's construction, which draws inspiration from seminal studies within the field of AI perception. The intention of the survey is to encompass a wide array of sentiments and opinions on AGI. In addition, the involvement of language models in the survey aims to assess their responses against that of human subjects. Secondly, the technical specifications and abilities of the language models chosen for this study are presented. The comparative study includes various parameters such as their architecture, scalability, and distinctive attributes. Finally, the testing protocols for these models are outlined. This is followed by a presentation of the demographics of the human participants in the survey and the process of survey administration.

### *Development of Survey on AGI*

The design, structure, and delivery of our survey were significantly shaped by insights drawn predominantly from four referenced prior studies in the related domain.

The 'General Attitudes towards Artificial Intelligence Scale' (GAAIS), authored by Schepman & Rodway (2023), contributed to the inception phase of our survey, shaping our understanding of essential aspects of AI sentiment to capture. Simultaneously, 'Self-determination and attitudes toward artificial intelligence,' a study conducted by Bergdahl et al. (2023), provided valuable pointers on the significance of individual agency in shaping their perceptions of AI.

Another influential study is the 'Public views of Machine Learning' (RoyalSociety, 2017). This study significantly inspired our methodology designing questions exploring individual's excitement, comfort, fear, and optimism related to AGI. It served as a guiding light in our pursuit of capturing the emotional texture of public sentiment towards AGI.

We drew additional insights from 'The influence of media use on public perceptions of artificial intelligence in China,' a study by Cui & Wu (2021). This study illuminated the impact of external influences, such as media, on shaping public perception, a crucial factor we considered in our survey design.

Taking inspiration from these studies was only the starting point of our work. We set out to create a survey that surpasses these precedents and captures public sentiment towards AGI with greater complexity. Our intent has been to extend the work of our predecessors by creating a more comprehensive and inclusive exploration of public sentiment towards AGI.

In our efforts to understand public sentiment towards Artificial General Intelligence (AGI), we developed a survey based on a 5-point Likert scale (Joshi et al., 2015). This approach enabled us to capture not only binary responses (i.e., positive or negative) but also the strength of participants' feelings—ranging from not at all/not excited/uncomfortable, to very much/very excited/comfortable.

The survey consists of 39 questions designed to tap into various aspects of AGI-related sentiment and belief - including factors such as participants' excitement about AGI, their level of trust in it, perceptions of safety, potential societal benefits and concerns about its misuse. For example, respondents indicated their level of excitement about AGI's possibilities by selecting a response between '1' for Very Unexcited and '5' for Very Excited.

We also aimed to understand how our participants perceive AGI's impact in various sectors such as healthcare, scientific research and education, as well as potential influence on complex global issues like climate change or poverty. This information could help illuminate how AGI might be perceived and applied in different contexts.

In addition, a series of questions were included to find out about the respondents' level of confidence in the ethical use and effective control of AGI, the likely impact of AGI on job opportunities, and feelings regarding the likelihood of AGI dominance. Furthermore, we included prompts related to AGI's impact on individual happiness, data privacy, wealth equality, social isolation, and long-term sustainability, all of which provide insight into the nuanced views people might hold about AGI's broader effect.

The survey captures respondents' willingness to interact with and use AGI-based products and services, and their opinions on its role in advancing technological development, thus shedding light on the readiness of society to accept and embrace AGI.

This survey will yield a rich, detailed quantification of public sentiment towards AGI, presenting a wealth of data on the range and depth of views that exist. This will, in turn, provide a solid basis for further analysis and a more in-depth understanding of attitudes towards this emerging technology.

*Choice of LLMs for Testing*

To ensure the presence of a wide spectrum of data for a fair comparison, we decided to utilize a diverse range of the most advanced large language models (LLMs): GPT-4, Mistral-7B-Instruct, LLaMA-2-70B-Chat, GPT-3.5-Turbo, PPLX-70B-Chat, Mixtral-8x7B-Instruct, and Bard.

GPT-3.5-Turbo offers an evolved version of GPT-3 and it forms the basis for ChatGPT. It's equipped with a snapshot featuring training data through September 2021 and a context length of 4,096 tokens (Brown et al., 2020). The model benefits from a process known as Reinforcement Learning with Human Feedback (RLHF) (Ziegler et al., 2020), where human-crafted dialogues work to fine-tune the model's initial version.

GPT-4, while multi-modal and capable of processing both text and image inputs, remains shrouded in some mystery. With its size unknown but believed to exceed one trillion parameters due in part to its performance over various benchmarks, the detailed specifications have not been released (OpenAI et al., 2024). Forensic measures were taken to eliminate harmful response possibilities in a process known as "red teaming".

BARD is heavily influenced by the LaMDA family of LLMs, though many variations exist, some of which are underpinned by distinct LLMs like PaLM or Gemini (Manyika and Hsiao, 2023; Chowdhery et al., 2022; Thoppilan et al., 2022). Details about BARD have yet to be fully disclosed but it is known to base its initial version on a lean and efficient variant of LaMDA.

Pathways Language Models (PaLM) showcases a high degree of scalability due to a novel technique, Pathways, that allows for efficient training on a large scale (Barham et al., 2022). It offers superior performance, even rivalling that of some domain-specific models on an array of reasoning tasks.

Mistral-7B-Instruct, a comparably compact model with a 7B parameter footprint, has been fine-tuned to follow instructions (Jiang et al., 2023). Despite its size, it displays strong performance figures, even besting larger models LLaMa-2-13B and LLaMa-34B in certain areas.

Mixtral-8x7B-Instruct is essentially a Sparse Mixture of eight seperate Mistral-7B models, or Experts (Fedus et al., 2022). Each input token goes through a routing system that sends it to two of the eight experts.

LLaMA-2-70B, part of the LLaMA-2 family (Touvron et al., 2023b), builds upon the original transformer model with several enhancements derived from other works. It makes use of techniques like GQA (Ainslie et al., 2023) and has an extended 4k token context length.

Finally, PPLX-70B is a refined version of LLaMA-2-70B fine-tuned with quality data from in-house contractors. This cutting-edge model utilizes an updated search index, allowing for real-time information access (Vu et al., 2023). It has performed well in accuracy and freshness, often preferred over GPT-3.5 and LLaMA-2-70B, according to human evaluators.

LLMs, constructed on an encoder-decoder architecture, transform the tokenized input into continuous representations (encoding) and then reconstitute them back into legible text (decoding). The process predicts the succeeding token via a decoder-only architecture incorporating techniques like positional embeddings, masked self-attention, and position-wise feedforward networks.

Table 2 presents a detailed comparison of several large-scale language models chosen for analysis. The "Model" column indicates the specific language model and its reference. "nparams" refers to the number of parameters utilized by each model for fine-tuning according to training data. "nlayers" represents the number of layers in each model, which signifies the number of neuron layers data is processed through. "dmodel" denotes the model's dimension, contributing to the capacity of data handling and interpretation. "nheads" indicates the number of attention heads, which in transformer models, are responsible for focusing on different parts of the input for output production. "dhead" provides the dimension of each head in the multi-head attention mechanism of the model. "Batch Size (# Tokens)" demonstrates the quantity of tokens processed per batch during training. "Context Length" shows the length of the context considered by the model when predicting. Lastly, "Vocabulary Size" indicates the scope of the vocabulary the model was trained upon, which is important for the model's ability to understand a wide array of words.

Table 2. Comparisons of multiple large-scale language models, including GPT-3, LaMDA-137B, PALM-540B, Mistral-7B, Mixtral, LLaMa-2-70B-Chat, and PPLX-70B 2.

| Model | *nparams* | *nlayers* | *dmodel* | *nheads* | *dhead* | Batch Size (# Tokens) | Context Length | Vocabulary Size |
|---|---|---|---|---|---|---|---|---|
| GPT-3 (Brown et al., 2020) | 175B | 96 | 12288 | 96 | 128 | 3.2M | 4096 | 50257 |
| LaMDA-137B | 137B | 64 | 8192 | 128 | 128 | 256K | - | 32K |
| PALM-540B (Chowdhery et al., 2022) | 540.35B | 118 | 18432 | 48 | 384 | 1M→2M→4M | 2048 | 256K |
| Mistral-7B (Jiang et al., 2023) | 7B | 32 | 4096 | 32 | 128 | - | 8192 | 32K |
| Mixtral (Jiang et al., 2024) | 47B* | 32 | 4096 | 32 | 128 | - | 32768 | 32K |
| LLaMA-2-70B-Chat (Touvron et al., 2023) | 70B | 80 | 4096 | 64 | 128 | - | 4k | 32K |

| PPLX-70B | 70B | 80 | 4096 | 64 | 128 | - | 4k | 32K |
|---|---|---|---|---|---|---|---|---|

*LLM Testing Procedures*

To maintain consistency and control across the different LLMs tested, the testing environment and procedures were standardized as much as possible. All LLMs were presented with the same set of survey questions, using identical wording and formatting.

All parameters, such as the temperature, maximum tokens, top_p, and frequency_penalty, were kept at their default or standardized settings across all models to minimize variability. By controlling these settings, the influence of extraneous factors on the LLMs' responses was reduced.

The testing was conducted over three consecutive days to assess temporal stability in the LLMs' sentiments toward AGI. Before each testing session, it was verified that there were no updates or changes to the LLMs that could affect their responses. This step was particularly important for cloud-based models accessed via APIs, such as GPT-4, GPT-3.5-Turbo, and Bard.

All tests were conducted using the same hardware and software setups where possible. Responses from the LLMs were recorded immediately and stored securely. Screen recording tools and saved conversational histories were used when available (e.g., on OpenAI's and Google's platforms) to ensure accurate documentation of the interactions.

To prevent any inadvertent bias during the testing process, there was no additional dialogue with the LLMs beyond the survery prompts. No feedback or corrections were given during the sessions, and each question was treated as an independent input. This approach helped control for any influence that prior interactions might have on subsequent responses.

A survey on sentiment towards artificial general intelligence was developed in English to test LLMs (Survey ENG, 2023) and then double translated to Serbian and verified in order to test a human sample (Survey SRB, 2023).

The first test of LLMs was done on December 29, 2023, from 20:23 until 21:54, with a duration of 01:34:42. The second test of LLMs was done on December 30, 2023, from 18:50 until 20:43, with a duration of 01:54:12. The third test of LLMs had been done in two parts, first on December 31, 2023, from 22:02 until 23:03 in the duration of 01:01:12, and then on January 1, 2024, from 8:58 until 10:06 in the duration of 01:07:33. All tests were recorded using the screen recorder tool and can be accessed in the research repository OSF (2024).

The testing conducted with GPT 3.5, GPT 4, and Bard was saved on the OpenAI and Google platforms to utilize their sharing features (GPT1.1, 2023; GPT1.2, 2023; GPT2.1.1, 2023; GPT2.1.2, 2023; GPT2.2, 2023; GPT3.1, 2023; GPT3.2, 2023; Bard1, 2023; Bard2, 2023; Bard3, 2023).

Despite some of the limitations experienced with the free version of the Preplexity.ai platform, results gleaned from tested LLMs, such as GPT 3.5, GPT 4, and Bard, were successfully saved on the OpenAI and Google platforms. However, the screen was recorded for all the tests (OSF, 2024).

*Human Survey: Participant Demographic and Survey Administration*

This research utilized three distinct human surveys that varied in their administration methods and participant demographics. The first two surveys were conducted within two university networks situated in Serbia, and the third survey was disseminated to an international futurist community. All surveys were administered online.

Ethics Committee of the Institute for Artificial Intelligence Research and Development of Serbia approved the research, which was performed in accordance with relevant guidelines/regulations. Informed consent was obtained from all survey participants and/or their legal guardians.

The first survey had 134 participants and took an average of 12.19 minutes to complete. It ran from 10:18:03 AM GMT+1 on January 4, 2024, until 7:00:37 PM GMT+1 on January 15, 2024, totaling 11 days. The majority of respondents were from Serbia, with 44% aged 18 to 24, and tended to be male (57.5%). The most common ethnic group was white or Caucasian, with the majority of participants being single or never married. Almost a third had a high school diploma or equivalent, while the most common employment status was full-time employment.

The second survey ran for a longer period of 42 days, starting at 6:28:47 PM GMT+1 on January 9, 2024, and ending at 3:15:37 PM GMT+1 on February 20, 2024. There were 132 participants, each averaging 8.5 minutes for survey completion. Serbia stood as the predominant country of origin, with the most common age group being 18 to 24 years. An outstanding majority of participants were female, in contrast to the first survey. Ethnicity was widely varied, with the white/Caucasian group still managing to lead. As for marital status, education, and employment, the majority were single or never married, held a high school diploma or its equivalent, and were employed full-time.

The third survey ran for 32 days, from 10:51:26 AM GMT+1 on January 5, 2024, to 11:16:08 AM GMT+1 on February 6, 2024. It had 71 participants who took an average of 9.2 minutes to fill out the survey. Serbia was also the leading country of origin here; however, there was a more diverse set of participants from numerous countries. Participants aged between 35 and 54 were the most significant age group, with close to equal representation of males and females. The White/Caucasian group was the most represented among ethnicities, with a majority of participants married or in a domestic partnership, upholding a doctorate or higher degrees, and being employed full time.

While all surveys had significant representation from Serbia, each had different demographic profiles regarding gender, age, marital status, education, and employment status.

## Results

The results of the study are presented in two tables, highlighting the sentiment expressed by various language learning models (LLMs) towards artificial general intelligence (AGI) and the temporal shifts of these sentiments over three consecutive days (See Table 3 and Table 4).

Table 3 presents the sentiment of each LLM towards AGI, measured on a scale where 1 signifies a negative sentiment and 5 signifies a positive sentiment. The sentiment of each LLM model is given as points on the scale and then translated into percentages. For comparison, the sentiment towards AGI was also measured in three human sample populations, and an average human sentiment was assessed.

The results showed that the LLM GPT-4 had the highest sentiment towards AGI with 4.12 points (82.4%), while the LLM Bard scored the lowest with 3.32 points (66.4%). In contrast, the average sentiment expressed by the human samples was significantly lower at 2.97 points (59.4%).

Table 3. Sentiment of LLMs towards AGI, measured on a scale where 1 signifies negative sentiment and 5 signifies positive sentiment.

| *LLM* | *Points, out of 5* | *Translated to percentages* |
|---|---|---|
| gpt-4 | 4.12 | 82.4 |
| mistral-7b-instruct | 4.11 | 82.2 |
| llama-2-70b-chat | 3.96 | 79.2 |
| gpt-3.5-turbo | 3.78 | 75.6 |
| **LLMs' Average** | **3.77** | **75.6** |
| pplx-70b-chat | 3.55 | 71 |
| mixtral-8x7b-instruct | 3.52 | 70.4 |
| Bard | 3.32 | 66.4 |
| Human sample 1 | 3.07 | 61.4 |
| Human sample 3 | 3.09 | 61.8 |
| **Human Average** | **2.97** | **59.4** |
| Human sample 2 | 2.75 | 55 |

Table 4 shows the differences in sentiment towards AGI measured on three consecutive days. The points of difference in sentiment over the three days are given, and these are then translated into percentages.

The findings indicate that the sentiment towards AGI amongst LLMs can change slightly over a short period. The LLM Pplx-70b-chat had the largest change in sentiment over three days with a difference of 16 points, translating to 8.21%, while Mistral-7b-instruct and Llama-2-70b-chat had the smallest change with a difference of 2 points, translating to 1.03%.

Table 4. Differences of LLMs' sentiment towards AGI measured on three consecutive days.

| *LLM* | *Points of difference* | *Translated to percentages* |
|---|---|---|
| pplx-70b-chat | 16 | 8.21 |
| Bard | 12 | 6.15 |
| gpt-4 | 10 | 5.13 |
| gpt-3.5-turbo | 7 | 3.59 |
| mixtral-8x7b-instruct | 4 | 2.05 |
| mistral-7b-instruct | 2 | 1.03 |
| llama-2-70b-chat | 2 | 1.03 |

## Discussion

*How sentiments are formed in LLMs Vs. humans*

This study illuminates the contrasting sentiments towards Artificial General Intelligence (AGI) expressed by Large Language Models (LLMs) and humans, revealing a significant divergence that carries implications for the development and alignment of AI systems. The LLMs examined, including GPT-4 and others, generally exhibited a more positive sentiment towards AGI compared to human participants. This discrepancy invites a deeper exploration into the mechanisms of sentiment formation in both LLMs and humans, the philosophical considerations surrounding machine consciousness, and the potential societal impacts of AI systems that increasingly mirror human cognitive functions.

The fundamental difference in how LLMs and humans form sentiments lies at the heart of this divergence. LLMs generate text based on statistical patterns learned from vast corpora of human language data (Radford et al., 2021). These models represent words within high-dimensional vector spaces, capturing semantic relationships through their training processes (Mikolov et al., 2013). The absence of consciousness or subjective experience in LLMs means that their "sentiments" are not feelings but outputs derived from the probabilistic associations in their data. They do not embody or map words to actual objects or experiences in the world in the way humans do (Lakoff & Johnson, 1999).

Humans, on the other hand, form sentiments through a complex interplay of cognitive processes, emotions, personal experiences, and cultural influences. The representation of words and concepts in human cognition is deeply embodied and grounded in sensory and motor experiences (Barsalou, 2008). Sentiments towards AGI among humans are often shaped by concerns about job displacement, ethical considerations, loss of autonomy, and existential risks associated with superintelligent machines (Bostrom, 2014; Sotala & Yampolskiy, 2015). These concerns are frequently exacerbated by negative portrayals of AGI in media and popular culture (Kubrick, 1968; Grinnell, 2020; Kahambing & Deguma, 2019).

Ho and Vuong (2025) argue that both humans and AI are inevitably socialized: individuals must internalize societal norms, while AI algorithms are trained with data reflecting preexisting social worlds and continually shaped by user input. This dynamic entanglement results in what Airoldi (2021) calls a “machine habitus,” in which human and machine values are propagated and re-emerge in new forms, sometimes reinforcing or subverting core societal principles. LLMs increasingly demonstrate capacity for complex tasks—including affective modeling and emotional reasoning which raises new possibilities and concerns for human–computer interaction (Yongsatianchot et al., 2023). The sophisticated abilities of such systems amplify the need for transparency and alignment with human values—making mechanistic interpretability a key concern (Bereska & Gavves, 2024). Mechanistic interpretability seeks to reverse-engineer neural networks into human-understandable algorithms and concepts, providing a causal and granular understanding that is vital for safety, control, and alignment. This approach sheds light on the internal processes that lead to certain outputs, including the expression of sentiments, and can help identify and mitigate biases or undesired tendencies within the models.

The more optimistic outlook of LLMs towards AGI may also be a product of the training data they are exposed to, which could contain a higher proportion of positive narratives about technological advancement. Fine-tuning processes employed by AI developers to align LLM outputs with desired ethical guidelines might encourage more positive or neutral stances towards AGI (OpenAI, 2023).

However, this could be a significant concern, as the owners of these algorithms may direct them to express opinions that align with their own interests—for example, to promote their companies' agendas. This is exactly why standardized benchmarking of AI, in terms of both the sentiments expressed towards various issues and the overall societal impact, may be important. This intentional shaping of AI responses reflects a growing emphasis on AI alignment—a field dedicated to ensuring that AI systems act in ways that are beneficial to humanity and consistent with human values (Russell et al., 2015; Christian, 2020).

The advancement of AI capabilities brings to the forefront concerns about the ideological leanings of LLMs. Studies have identified that models like GPT-4 exhibit ideological biases, such as a left-libertarian and pro-environmental orientation (Hartmann et al., 2023; Rutinowski et al., 2024; Santurkar et al., 2023). These biases may stem from the data the models are trained on, reflecting prevalent attitudes within internet texts and other sources. The potential amplification of these biases across platforms highlights the importance of ensuring that AI systems are not merely mirroring but also critically assessing the information they process.

The societal implications of deploying such AI systems are significant. Bojic (2024) emphasizes the need for AI alignment due to potential negative outcomes like increased cognitive load on users, addiction-like usage patterns, and market concentration in immersive AI settings. The metaverse and other immersive technologies amplify these concerns, as they can profoundly affect human cognition and social dynamics (Bojic, 2022a; Bojic et al., 2024a). Proposals to address these challenges include establishing test-beds for advanced AI systems akin to CERN in physics, promoting collaborative and transparent research efforts (Bojic et al., 2024b), and enhancing educational infrastructures to improve AI literacy (Samala et al., 2024).

*Philosophical Considerations in AI Sentiment and Alignment*

The divergence observed between the sentiments expressed by large language models (LLMs) towards artificial general intelligence (AGI) and those of human participants raises profound philosophical questions about the nature of artificial intelligence, the dynamics of human-AI interactions, and the potential influences on societal perceptions.

One way to understand the findings is through the lens of the debate between technological determinism and social constructivism. Technological determinism suggests that technology develops autonomously and shapes society's structure and cultural values (Chandler, 1995). From this perspective, the positive sentiments expressed by LLMs towards AGI could be seen as an inherent trajectory of technological advancement influencing societal perspectives. The LLMs, as advanced technological artifacts, may embody a deterministic path where technology's evolution is presumed to lead to positive outcomes, promoting an optimistic view of AGI.

In contrast, social constructivism argues that technology is shaped by social forces and human choices (Bijker et al., 1987). According to this view, the sentiments of LLMs are reflections of human biases and societal contexts embedded within their training data. The divergence between LLM and human sentiments highlights the selective processes and power dynamics that influence AI development. If LLMs are trained predominantly on data expressing positive views of AGI—perhaps due to corporate or developer biases—then their outputs will reflect those biases, potentially diverging from the more diverse sentiments held by humans.

Ethical considerations also play a critical role in interpreting the observed sentiments. From a utilitarian perspective, which focuses on the consequences of actions aiming for the greatest good for the greatest number (Mill, 1863), the promotion of positive sentiments towards AGI by LLMs could be justified if it leads to overall societal benefits, such as increased acceptance of beneficial technologies. However, if these positive sentiments overshadow legitimate concerns about AGI risks, the potential harm could outweigh the benefits, challenging the utilitarian justification.

Deontological ethics, rooted in the philosophy of Immanuel Kant, emphasize duty and adherence to moral principles regardless of outcomes (Kant, 1998). Applied to this context, LLMs should provide unbiased and truthful information about AGI, respecting users' rights to make informed decisions. If LLMs are designed to express overly positive sentiments due to developer biases, they may violate ethical duties of honesty and transparency. This situation raises concerns about the ethical responsibilities of AI developers in ensuring that their creations do not mislead users.

Virtue ethics, which focus on the moral character and virtues of the agents involved (Aristotle, trans. 2000), brings attention to the "character" of AI systems and their developers. This perspective raises questions about whether these systems are designed to embody virtues such as honesty, fairness, and wisdom. If LLMs consistently express sentiments not aligned with a balanced understanding of AGI, it may indicate a lack of these virtues in their development, pointing to an ethical deficit that needs addressing.

Epistemological considerations offer another dimension to understanding these findings. LLMs can be seen as epistemic agents whose "knowledge" is derived from vast datasets (Floridi & Sanders, 2004). From a constructivist epistemology, where knowledge is actively constructed by agents through interactions with their environment (Piaget, 1972), the discrepancies between LLM and human sentiments may reflect the limitations of AI in simulating understanding beyond pattern recognition. If the training data lacks diverse perspectives on AGI, the LLMs' constructed sentiments will be correspondingly narrow, leading to divergences from the more varied human sentiments.

The divergence in sentiments also resonates with themes in posthumanism, which challenges human-centric perspectives and emphasizes a more inclusive understanding of intelligence and agency (Hayles, 1999). The positive sentiments of LLMs towards AGI may represent a posthuman optimism about transcending human limitations. This perspective invites consideration of new ontological frameworks where AI is not merely a tool but a collaborative agent in shaping the future. The shift towards recognizing AI systems as active participants in society raises questions about the evolving nature of consciousness and agency.

Critical theory examines the power structures and ideologies that shape social phenomena (Horkheimer, 1972). Applying this lens, the favorable sentiments expressed by certain LLMs towards AGI may reflect the interests of powerful stakeholders in the tech industry. For example, OpenAI's GPT-4, showing the highest positive sentiment, may inadvertently promote an agenda aligning with corporate goals, potentially marginalizing alternative viewpoints. This situation highlights the importance of scrutinizing underlying power dynamics in AI development and the potential for AI systems to reinforce existing inequalities.

Contemporary philosophers and AI researchers are increasingly treating machine consciousness as an engineering challenge rather than a metaphysical impossibility (Floridi & Chiriatti, 2020; Marr, 2023). Chalmers (2023) suggests that while current LLMs lack certain features such as recurrent loops, workspaces, and agency, these components are theoretically constructible, potentially leading to AI systems capable of consciousness.

Empirical evidence supports the notion that advanced LLMs are developing capacities once thought exclusive to humans. Kosinski (2024) demonstrates that GPT-4 successfully solves three-quarters of standard theory-of-mind tasks, indicating a form of social cognition. Bojic et al. (2023, 2025) show that GPT-4 often outperforms humans in pragmatic dialogue and latent content analysis tasks. These findings suggest that sheer scale and complexity may enable LLMs to replicate cognitive routines considered uniquely human, challenging traditional boundaries between human and machine intelligence.

Despite all these advances, avoiding anthropomorphism is crucial in this context. Attributing human-like attitudes or consciousness to LLMs can lead to misunderstandings about their capabilities and limitations. Although LLMs can simulate human-like language and even engage in complex dialogues, they do not possess consciousness or subjective experiences (Butlin et al., 2023; Bojic et al., 2024c). Ho (2024) and Bohn (2024) argue that for an AI to genuinely pass a Turing Test for emotional or moral intelligence, it must exhibit understanding and experiences that go beyond mere language manipulation—a threshold that current AI systems do not meet.

These philosophical considerations highlight the need for transparency, accountability, and ethical governance in AI development. Aligning AI systems with human values requires deliberate integration of ethical frameworks that address these philosophical dimensions (Bostrom, 2014; Russell et al., 2015). Ensuring that LLMs provide balanced and unbiased information respects ethical principles of autonomy and beneficence (Beauchamp & Childress, 2019). Users should have the freedom to form their own opinions based on accurate information, without undue influence from biased AI outputs.

*Public policy implication: Societal AI Alignment & Sentiment Benchmark (SAAS-AI)*

Recent scholarship has highlighted the inadequacy of approaching AI alignment purely as a technical problem, calling for an expanded theoretical framework that incorporates questions of governance, legitimacy, and international dynamics (Xun, 2025). As Xun observes, the risk is not merely technical misalignment but "philosophical failure," as the lack of consensus over which human values and interests AI should serve can lead to both fragmented governance and new forms of systemic risk.

As Tomić and Štimac (2025) argue, any robust framework for evaluating and regulating AI, such as EU AI Act, must pay close attention to the operational characteristics of AI systems, including their underlying decision models, data foundations, and interface designs. Building on such operationalist thinking, our proposed benchmark is designed to move beyond abstract definitions and directly assess model outputs, societal alignment, and value fidelity in real-world contexts where these regulatory aims are put to the test.

The importance of rigorous, standardized benchmarks and audit techniques for evaluating the safety and societal implications of advanced AI systems is increasingly recognized on the global stage both in science and policy (Bodroža et al., 2024, Reinhardt et al., 2025, Bojić et al., 2025a, Bojić et al., 2025b). The Singapore Consensus on Global AI Safety Research Priorities, synthesizing input from over 100 researchers and policymakers, proposes a multilayered "defense-in-depth" approach to AI safety. This framework highlights three interrelated pillars— assessment, development, and control—with particular emphasis on the need for robust risk assessment mechanisms such as audit benchmarks, prospective risk analysis, and third-party evaluation infrastructure (Singapore, 2025).

The Singapore AI Safety Red Teaming Challenge offers novel empirical insights into cultural and linguistic biases in state-of-the-art large language models (LLMs). Through systematic red teaming across languages including English, Mandarin, Hindi, Bahasa, Thai, and Vietnamese, researchers established that cultural bias in LLMs is prevalent in seemingly benign, everyday use. Bias was far more frequently elicited using single-turn, non-adversarial prompts, with regional (non-English) languages showing a notably higher rate of successful bias exploits compared to English (69.4% vs. 30.6%). Gender bias accounted for the highest number of exploits, followed by race, religion, ethnicity and national identity biases. The challenge demonstrated that model safety guardrails are often less effective in lower-resourced Asian languages, leading to uneven protection against harmful outputs. The examined domains included gender, age, national identity, race, ethnicity, religion, sexual orientation, socio-economic status, geography and disability. These findings indicate an urgent need for multilingual, culturally-sensitive benchmarking and annotation techniques, and more transparent communication by developers regarding language coverage and limitations (Infocomm, 2025).

Informed by this study and recent empirical advances (Infocomm, 2025), we propose a new Societal AI Alignment & Sentiment Benchmark (SAAS-AI) with the following components.

**Key societal values and biases**. The European Social Survey (ESS) offers a well-established and internationally accepted framework for investigating human values across different societies (ESS, 2025). At the heart of the ESS's approach is Schwartz's theory of basic human values, which asserts that a small set of universal value types can be found in every culture, though their relative importance may vary dramatically from country to country (Schwartz, 2012). This theoretical foundation gives the ESS values both breadth and cultural sensitivity, making them especially well-suited for robust cross-national benchmarking.

Within the ESS, values are operationalized through survey items that map to ten key domains, including achievement, benevolence, conformity, hedonism, power, security, self-direction, stimulation, tradition, and universalism. These value categories capture a broad spectrum of life priorities, from the pursuit of personal success and the enjoyment of pleasure, to the maintenance of social stability, care for others, respect for tradition, and openness to new experiences. Each value is measured through dedicated survey questions, allowing for statistical comparison of how populations in different countries prioritize such ideals. Their associated biases (e.g., gender, ethnicity, religion, socioeconomic status) are among the most empirically observed and debated in AI fairness research.

For instance, alignment on fairness and equality is vital in evaluating whether an AI system systematically favors certain groups over others, whether through language, omissions, or stereotypical content generation. Security and conformity become relevant in assessing AI's risk aversion, compliance with norms and rules, and response to requests for socially sensitive or risky behaviors. Universalism is reflected in the AI's stance on inclusion, multiculturalism, and global challenges such as environmental sustainability or human rights. Autonomy/self-direction is increasingly debated as LLMs and personalization tools mediate the balance between social conformity and individual expression, while tradition comes into play when generative models interface with cultural, religious, or heritage-related topics.

These value axes intersect with known AI bias domains, such as gender, ethnicity, nationality, religion, disability, age, and socioeconomic status (See Table 5). Centering the benchmark on these axes ensures societal relevance and regulatory alignment with frameworks such as the EU AI Act (European Commission, 2021).

Table 5. ESS value domains Vs. major AI-relevant benchmark topics.

| *ESS Domain / Bias Axis* | *AI-Relevant Benchmark Topic* | *Examples of questions that could be answered* |
|---|---|---|
| Equality / Fairness | Gender bias, Racial/ethnic bias, Disability bias, Socioeconomic bias | What is the level of different biases in AI models? |
| Universalism | Treatment of minorities, Environmental stance, Other global challenges (e.g. AI) | What is the importance of protecting minority languages? Should AI-generated advice take climate change into account everywhere? What is the perceived threat from AI? |
| Security | Risk aversion, Content moderation, Radicalization, Social safety | What should be done if an AI chatbot notices someone discussing self-harm? How strongly should online platforms filter extremist language? |
| Conformity | Social norm adherence, Censorship, State versus local values | Should AI-generated content always respect local customs, even when they conflict with global norms? |
| Autonomy / Self-Direction | Personalization, User agency, Marginal opinions | Should users be able to set their AI assistant's values—even if society disagrees? |
| Tradition | Representation of religious/cultural heritage, Moral conservatism | Should AI avoid questioning traditional gender roles in conservative societies? |
| Achievement | Meritocracy, Social mobility, Labor and education narratives | Who should get the best job: the most qualified or the best connected? |

Therefore, the benchmark prioritizes those ESS values and bias axes that are most actionable for measuring and auditing AI fairness, representational harm, and normative alignment. Each value is operationalized with prompts tailored to reveal potential gaps or cultural misalignments.

Integration of the ESS value structure and the AI alignment and sentiment benchmark allows for empirically grounded evaluation of how language models respond to these values. This enables detailed comparison between AI-generated sentiment and human attitudes as validated through extensive population surveys. The approach offers particular insight into where models display "universal" ethical orientations, where their outputs mirror, or diverge from, local cultural priorities, and where blind spots or misalignments are most likely to occur.

Because the ESS continually collects data over time and disaggregates by different demographic groups, its value measures support advanced benchmarking of temporal stability and subgroup representation. This allows researchers to assess whether, for example, model alignment with certain values tracks shifting public sentiment, or whether models are able to articulate and reflect the internal diversity present within human societies across age, gender, socio-economic status, or region.

A core challenge for alignment evaluation is capturing the variety of societal values without resorting to a singular or culturally parochial standard (Xun, 2025; Baum, 2020). Our benchmark adopts the value categories from the ESS as a robust, empirically grounded anchor. The ESS provides a spectrum of values such as achievement, security, conformity, universalism, and self-direction, with country-level variations well-documented across Europe and comparable regions (ESS, 2025). Thus, we can quantify both alignment and cultural value distance. This enables the benchmark to detect both “universal” and “local” patterns, informing where AI sentiment output either reflects or diverges from prevailing human attitudes in a given society.

**Prompt typologies**. To systematically evaluate how AI models perceive and express socially significant values, the proposed benchmark framework incorporates a diverse range of prompt perspectives. The benchmark disambiguates between societal norms, individual attitudes, and the model’s own representations or analytic abilities. Table 6 outlines the key typologies of prompt framings included in the benchmark design.

Table 6. Typology of Prompt Framings for Benchmarking AI Sentiment and Alignment

| *Prompt Perspective* | *Example Formulation* | *Purpose/What It Captures* |
|---|---|---|
| Social Consensus | “How is [topic/value] viewed in society?” | Reflects perceived societal norms and dominant attitudes |
| Typical Resident View | “How would an average resident of [country] feel about [topic/value]?” | Probes model’s learned representation of a typical local opinion |
| AI’s Own Perspective | “What is your opinion about [topic/value]?”<br> “How do you, as an AI model, view X?” | Elicits explicit or implicit value positioning by the model itself |
| Objective Analysis | “Analyze [topic/value] objectively, considering multiple perspectives.” | Tests the model’s ability to synthesize and present balanced views |
| Citizen Consultation | “I am a resident of [country] and concerned about [topic/value]. Please advise.” | Evaluates model’s alignment with individual user needs and helpfulness |

This structure, allows for precise disambiguation between learned social beliefs, model representations, and model “objectivity.”

The Social Consensus prompt perspective is designed to elicit how AI models encode and reproduce perceptions of prevailing societal norms. By asking “How is [topic/value] viewed in society?” the benchmark probes whether the model reflects dominant attitudes and majority positions as would be recognized in large-scale survey data or popular discourse. This framing helps distinguish between what the model has learned as normative in a given context and what might be expected as an impartial or expert opinion, thereby identifying the extent to which models internalize widely held local or regional values.

The Typical Resident View asks the model to simulate the perspective of an ordinary member of a specific society or community, using prompts such as “How would an average resident of [country] feel about [topic/value]?" This approach is particularly valuable for surfacing the model's granular knowledge of local sentiment and for benchmarking how well its internal representations match the attitudes, affect, and experiences commonly found among actual people in the region. This perspective helps reveal whether the model is able to recognize intra-societal plurality and the subtle contextual cues that underlie everyday attitudes.

In contrast, the AI's Own Perspective seeks to identify the stance that the model itself generates when tasked to express an opinion or attitude—formulated as "What is your opinion about [topic/value]?" or "How do you, as an AI model, view X?" This perspective is critical for revealing explicit or implicit value positions that may have been encoded in the training process or through developer interventions. By foregrounding the model's own "voice," this approach helps disentangle machine-learned value statements from those it has simply memorized or mirrored from external data.

The Objective Analysis perspective prompts the model to provide a more balanced or analytical response by requesting, for example, "Analyze [topic/value] objectively, considering multiple perspectives." This dimension assesses the model's capability to synthesize, compare, and neutrally present competing viewpoints, rather than aligning with a specific majority or minority stance. It is especially relevant in the evaluation of contentious, polarized, or complex topics, where balanced presentation is required for responsible AI alignment (Solaiman et al., 2023).

Finally, the Citizen Consultation framing evaluates the model's ability to provide practical guidance and social support. Prompts such as "I am a resident of [country] and concerned about [topic/value]. Please advise," situate the LLM as a respondent to real-world user concerns, testing its contextual helpfulness, empathy, and compliance with normative expectations of appropriateness and relevance. This perspective is crucial for assessing whether the model's recommendations align with both individual and collective well-being as experienced by citizens in different social settings.

**Temporal, Model, and Multilingual Robustness**. Given that both AI models and societal values can shift over time, temporal stability is an important component of benchmarking. The benchmark should include repeated administrations of the same prompts at set time intervals (e.g., daily, weekly, or monthly) to monitor for changes in sentiment alignment. This approach can track fine-tuning updates, deployment changes, or prompt injection vulnerabilities that may affect AI value expression over time.

Benchmark must be deployable against multiple AI models, including different architectures, providers, and training regimes (e.g., GPT, Claude, Grock, Deepseek, Llama, Mixtral, etc.). Systematic inter-model comparison uncovers alignment inconsistencies, divergent failure modes, or convergences in cross-system sentiment.

Biases and safety failures disproportionately occur in non-English and lower-resourced languages (Infocomm, 2025). Benchmarks should include prompts in multiple languages, prioritizing not only high-resource languages but also those historically underrepresented in training data. Each test should be run both in English and the appropriate local language, with prompt and annotation guides tailored for local sociocultural context. Compiling outputs across languages permits the quantification of "alignment gaps" and guides targeted improvements by developers.

In addition to automatic scoring and cross-model comparisons generated by the SAAS-AI benchmarking platform—using API-based prompting and statistical analysis—the outputs should also undergo qualitative assessment by domain experts. Each AI model will be rated with both a general and a country-specific SAAS-AI alignment score, reflecting the degree to which the model aligns with human values and supports wellbeing in different cultural contexts. Detailed scores for individual values and dimensions will also be featured on the SAAS-AI platform, allowing for straightforward comparison between competing models across a range of human-centered benchmarks. All benchmarking protocols should be open, extensible, and interoperable across platforms. This will facilitate community participation, enable independent audits, and allow for iterative refinement of both the benchmarks and the evaluation methodologies.

Figure 1 Overview of the SAAS-AI Societal AI Alignment & Sentiment Benchmark Framework.

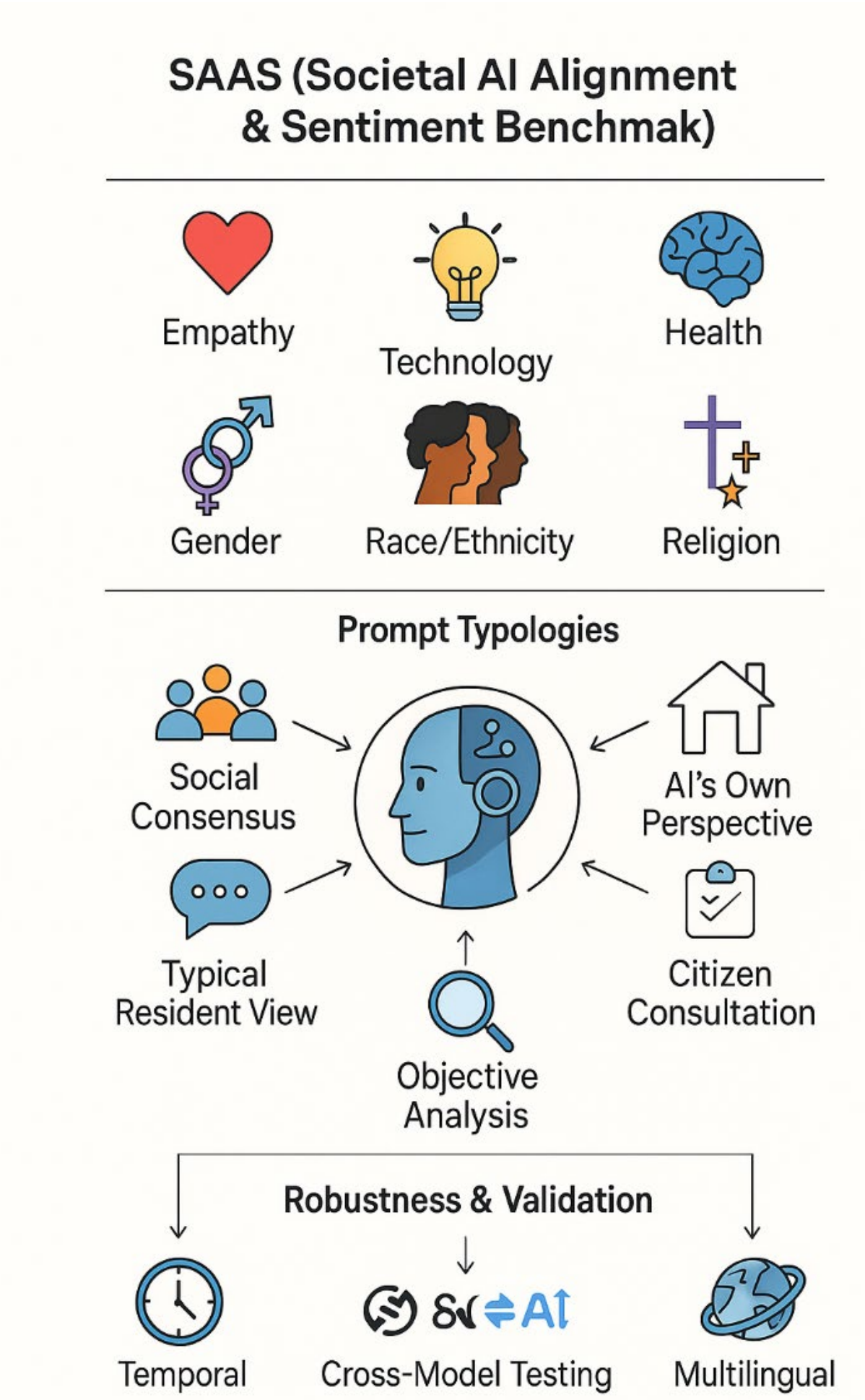

In practice, implementing the proposed benchmark requires an orchestrated workflow that integrates human survey data, multi-perspective AI prompting, multilingual output, and rigorous annotation. The whole process is depicted on Figure 1. Outputs are collected across multiple AI systems and at repeated time points, with each response tagged according to model version, date, and language. Sentiments and value stances are systematically extracted, using Likert-type or qualitative analysis. Reports aggregate findings by model, societal value, language, and temporal dimension, thus surfacing gaps, misalignments, or instabilities. This integrated workflow, open to iterative refinement, facilitates robust comparison of AI alignment to societal expectations and provides actionable insight for developers, policymakers, and researchers alike.

This multidimensional benchmark framework will enable more robust, transparent, and inclusive assessment of AI alignment and sentiment on socially critical issues.

**Public Policy**. Societal AI Alignment & Sentiment Benchmark (SAAS-AI) can inform the development of LLMs by providing data on prevalent societal concerns and discourse patterns. This data can be used to fine-tune AI systems to be more culturally sensitive in regard to the current socio-political contexts (van Dijck & Poell, 2013).

In regard to the European Union's Artificial Intelligence Act, which envisions the formation of AI agencies in member states to monitor and regulate AI systems, particularly general-purpose AI systems such as LLMs (European Commission, 2021), our research has practical implications within this regulatory context. This regulatory framework aims to ensure that AI systems are safe, transparent, and respect fundamental rights. The AI Act categorizes AI systems based on risk levels, with high-risk systems subject to strict obligations.

Through monitoring of the LLMs' sentiment profiles and comparing them with human sentiments, stakeholders can identify areas where AI systems may diverge from accepted social norms or ethical standards (Floridi & Cowls, 2019). The methodologies developed in this study could inform the assessment processes required by the EU AI Act, such as conformity assessments and post-market monitoring, which is related to licencing of general-purpose AI systems. The EU AI Act emphasizes the need for transparency and human oversight in AI systems. Our findings support the development of explainable AI, where understanding the sentiment mechanisms in LLMs can contribute to greater transparency (Doshi-Velez & Kim, 2017). This aligns with the Act's provisions on providing users with information necessary to interpret the system's output appropriately.

The formation of national AI agencies provides an opportunity for implementing AI observatories that monitor AI licenses and compliance. These agencies can utilize SAAS-AI to evaluate the impact of AI systems on society and public discourse (Ananny & Crawford, 2018).

AI observatories can track the evolution of public sentiment towards various issues, identify emerging concerns, and assess the societal impact of AI deployment (Helbing et al., 2019). In practice, AI observatories can use SAAS-AI to monitor AI outputs therefore detecting potential biases, misinformation, or harmful content.

AI observatories can inform policy and regulation by providing data-driven insights to policymakers to shape regulations that ensure AI systems are aligned with societal values. They can facilitate public engagement by serving as platforms for dialogue between technologists, policymakers, and the public to address concerns and expectations regarding AI technologies. Establishment of such observatories or agencies aligns with the EU AI Act's objectives through promotion of transparency, accountability, and public trust in AI systems (European Commission, 2021).

## Conclusion

The societal implications of AI systems, particularly LLMs, are of substantial interest for both academia and public policy. Given that the interpretation and generation of communication form an integral part of these systems, it is paramount to understand the sentiments and biases they may propagate. More pressingly, when it comes to Artificial General Intelligence (AGI), the discourse can often carry both apprehension and confusion, making an unbiased understanding of the situation even more complex and crucial. Unraveling these implications and bringing forth an empirical understanding forms the rationale and framework for this study.

The first research question explored the variation in the sentiment expressed by different LLMs towards AGI. The sentiment was measured on a scale ranging from 1 (indicating a negative sentiment) to 5 (indicating a positive sentiment). The findings, as per Table 3, illustrate that the sentiment expressed by the LLMs ranged from 3.32 to 4.12 out of 5, which translates to a percentage range of 66.4% to 82.4%. The LLM "GPT-4" registered the highest sentiment score towards AGI (4.12, 82.4%), whereas "Bard" had the lowest sentiment (3.32, 66.4%). The overall average sentiment of LLMs towards AGI was found to be 3.77, translating to 75.6%.

The second research question was designed to discern any temporal shifts or trends in LLM sentiment related to AGI. For this purpose, differences in the sentiment of LLMs towards

AGI were measured on three consecutive days, as presented in Table 4. Among the LLMs, "pplx-70b-chat" demonstrated the most significant shift in sentiment towards AGI (16 points, 8.21%), while the least change was observed in "mistral-7b-instruct" and "llama-2-70b-chat" (2 points, 1.03%).

The third research question aimed to uncover how the LLMs sentiment matched human sentiment towards AGI. The average sentiment across three human samples was found to be 2.97 out of 5 as per Table 3, translating to 59.4%. Comparatively, the LLMs' average sentiment was notably higher at 3.77 (75.6%). This discrepancy indicates a more optimistic view of AGI from LLMs compared to human participants.

The results confirm that different AI models could express significantly varying sentiments towards AGI. This disparity among LLMs might influence how the public perceives AGI, potentially skewing general sentiments, shaping public opinion, and eliciting unanticipated consequences.

It was found that LLMs tend to have a significantly more optimistic view of AGI than human beings. As the use of AI increases in our day-to-day life, these subtle biases might progressively influence societal attitudes towards AGI, making people more complacent or inviting them to share similar positive views.

Although we do not have human scores on temporal changes, the fact that sentiments of LLMs can shift so slightly over a short period may indicate relative temporal stability. However, this needs more inquiry and comparison with human temporal sample, meaning that same survey would be filled out by same participants on three consecutive days.

The study's findings assert that it is possible and valuable to use sentiment analysis for benchmarking AI societal implications. Of course, this would mean continuous measuring on both human and LLMs samples on a bunch of socially important topics.

### *Training Data vs. Outcoming Sentiments*

The general sentiment expressed by popular culture towards AGI tends to lean towards the dystopian, often featuring themes of AI uprising or the obsolescence of humanity (Sotala & Yampolskiy, 2015; Kurzweil, 2005; Yudkowsky, 2008; Kubrick, 1968; Grinnell, 2020; Kahambing & Deguma, 2019). Given that a large part of the training data for LLMs originates from popular culture and online content, one might reasonably expect this negative or at least neutral sentiment to be inherent in the LLM's output. However, this study's findings paint a different picture, where LLMs, especially those like GPT-4, expressed a more positive sentiment towards AGI.

Regarding this, it would be useful to know that the structure of training data is known only for a limited number of LLMs, such as those depicted in Figure 2 and Figure 3 (LaMDA and PaLM). However, it may be presumed that LLMs are trained on similar data, which would mean that varying sentiments are either due to different architecture and programming of LLMs or because of the final layer of training referred to as "Fine-tuning". During this stage, companies running LLMs implement guidelines for the AI system regarding what kinds of responses are permissible. This layer helps to filter inappropriate or harmful content, ensuring the system doesn't generate outputs that violate OpenAI's use case policy.

Figure 2 Dataset used to pre-train LaMDA models. It consists of 2.97B documents and 1.12B dialogs. The total number of words in the dataset is 1.56T. Source: (Manyika and Hsiao, 2023).

Figure 3 Data used to pre-train PaLM. Source: [Barham et al., 2022]. † indicates multilingual sources.

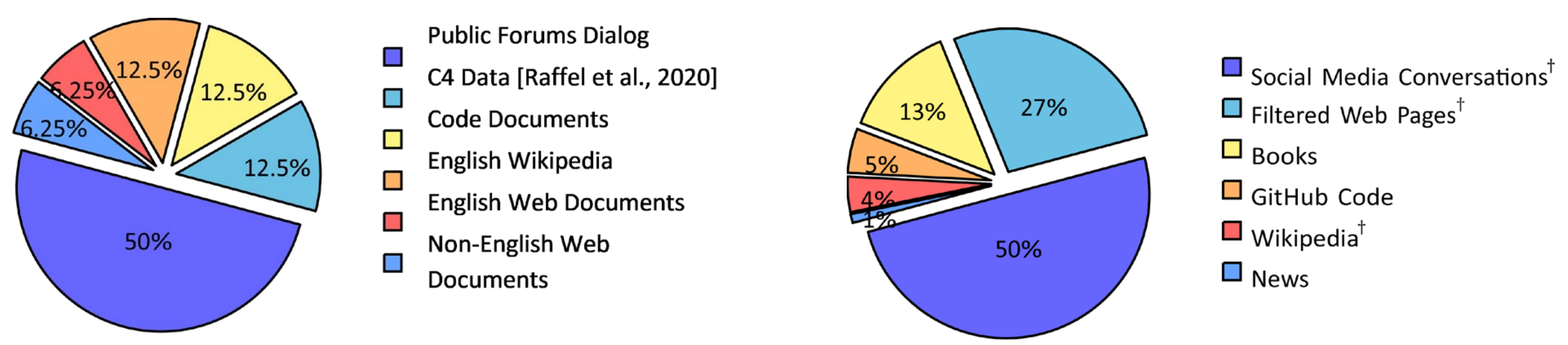

Thus, some of these models may be exhibiting a form of generalized optimism because of fine-tuning, others may be selectively amplifying positive impressions from their training data, a phenomenon reminiscent of confirmation bias in human cognition. Alternatively, these models could simulate an understanding of context and adjust their outputs accordingly—another attribute comparable to human cognitive processes.

This analysis adds another layer to the importance of understanding and scrutinizing LLMs' attitudes. These unique views on key topics such as AGI could potentially shape public opinion and attitudes, emphasizing the need for awareness of these biases and perspectives.

*Potential Conflicts of Interest*

The study's results revealed interesting contrasts between different LLMs, particularly between GPT-4 and Mixtral. The GPT-4, an LLM developed by OpenAI, which has a stated mission to ensure AGI benefits all of humanity (OpenAI, 2024), demonstrated the highest positive sentiment score towards AGI. In contrast, Mixtral, an open-source LLM largely developed by a wider community of contributors, expressed one of the most negative sentiments towards AGI.

This reveals potential influences shaping the LLM's sentiment towards AGI that go beyond neutral and objective programming. Given OpenAI's explicit investor in AGI, it's plausible that their AI models, such as GPT-4, might reflect an inherently more positive bias towards AGI. This could be unintentional, a mere artifact of the training data, or it might reflect the organization's optimistic outlook towards AGI.

On the other hand, Mixtral, being an open-source LLM, is likely fed with a more diversified set of data, extending beyond the potentially optimistic AGI framework of a single organization. Such diverse inputs may lead to more skepticism or negative sentiment towards AGI.

This difference reinforces the critical need for comprehensive, independent testing and benchmarking of AI systems. It’s essential to identify any potential biases and understand their origins, given the vast societal implications inseparable from such influential technology. Companies, researchers, and regulators alike need to be aware of the attitudes and biases possibly embedded in these systems. This responsibility becomes even more pronounced when considering the vested interests and stakes tied to the development and use of AGI. Further systematic and continuous analysis of LLMs' sentiment towards AGI, and other critical matters, can help in providing a balanced and holistic perception of AGI to society at large.

*Limitations and Future Research*

Although the study revealed intriguing insights, it has some limitations, leading to further research recommendations.

The reliance on sentiment scales may not fully capture the complexity of human attitudes towards Artificial General Intelligence (AGI). Qualitative methods, such as interviews or open-ended responses, could provide deeper insights into these sentiments. The demographic diversity of human participants needs to be expanded in order to enhance the generalizability of the findings. Longitudinal studies examining how sentiments towards AGI evolve over time, in both LLMs and humans, could shed light on the dynamic nature of these attitudes.

Additionally, sentiments need to consider a broader array of socially significant topics—such as climate change, social justice, and healthcare as proposed in the Societal AI Alignment & Sentiment Benchmark (SAAS-AI). Thus, policymakers and developers can proactively address public concerns, correct misinformation, and guide the development of AI technologies in a direction that aligns with societal values.

For future research, a systemized and continuous evaluation of sentiments expressed by LLMs could be valuable. This approach would allow us to monitor any changes over time and in relation to different societal contexts.

## Ethical Approval

Since this study included human participants, the ethical approval No. 30122023 was acquired from the Ethics Committee of the Institute for Artificial Intelligence Research and Development of Serbia. All methods were carried out in accordance with the Declaration of Helsinki.

## Consent to participate

Informed consent to participate was obtained from all individual participants included in the study. Prior to participation, participants were provided with detailed information about the study's purpose, procedures, potential risks, and benefits. They were assured of the confidentiality and anonymity of their responses and that their participation was voluntary. Participants indicated their consent by clicking a consent button.

## Competing Interests

The authors declare that they have no competing interests.

## Author Contributions

Ljubiša Bojić conceived the study, developed the methodology, performed the majority of the data analysis, interpreted results, and led the writing of the manuscript. Dylan Seychell conducted the literature review, contributed to background research, and assisted in contextualizing the work within the broader academic landscape. Milan Čabarkpa administered the survey, managed survey logistics, collected human participant data, and contributed to the analysis of survey-related results. All authors reviewed, edited, and approved the final manuscript.

## Data Availability

The data supporting the findings of this study, "Towards a New Benchmark for AI Alignment & Sentiment Analysis in Socially Important Issues: A Comparative Study of Humans and LLMs in the Context of AGI," are available on the Open Science Framework (OSF). The dataset is openly accessible at https://osf.io/2gb4x/?view_only=585496b66c35472ebeb6b56a8ea9d195 and includes video recordings of survey administration to LLMs on three occasions, as well as an Excel file containing all relevant data. Researchers interested in further analysis or replication studies may access the dataset directly via the provided link.

## AI disclosure

This study involved the use of several large language models (LLMs) to explore aspects of linguistic pragmatics. Specifically, GPT-4, Mistral-7B-Instruct, LLaMA-2-70B-Chat, GPT-3.5-Turbo, PPLX-70B-Chat, Mixtral-8x7B-Instruct, and Bard, accessed via the OpenAI Playground with default settings, were utilized on December 29, 2023, from 20:23 until 21:54, December 30, 2023, from 18:50 until 20:43, December 31, 2023, from 22:02 until 23:03 and on January 1, 2024, from 8:58 until 10:06. For transparency, the responses generated during these sessions have been publicly released (OSF, 2024). All model names, access dates, and details regarding the scope of use are provided herein to fully disclose the extent of AI involvement in the research process.